\documentclass[prl,twocolumn,superscriptaddress]{revtex4-1}

\usepackage{amsmath}
\usepackage{braket}
\usepackage{mathtools}

\usepackage{pgfplots}

\usepackage{hyperref}
\usepackage{lipsum}
\usepackage{xcolor}
\usepackage{comment}
\usepackage{bm,amssymb,graphicx,braket,dsfont,mathtools,bbold,MnSymbol}

\usepackage[latin1]{inputenc}
\usepackage[caption=false]{subfig}
\usepackage{pgfplots}
\usepackage{cancel}
\usepackage{tabularx,booktabs}
\usepackage{multirow}
\usepackage{fancyhdr}
\usepackage{amsthm}
\usepackage{inputenc}
\usepackage{soul}

\newcommand{\comm}[2]{\left[ #1, #2 \right]}

\newcommand{\lan}{\left\langle}
\newcommand{\ran}{\right\rangle}

\newcommand{\av}[1]{\lan #1 \ran}

\newcommand{\rme}[1]{{\rm{e}}^{#1}}

\newcommand{\Z}{\mathbb{Z}}

\newcommand{\lt}{\left(}
\newcommand{\rt}{\right)}

\begin{document}

\title{Singularities in large deviations of work in quantum quenches}

\author{P. Rotondo}
\affiliation{School of Physics and Astronomy, University of Nottingham, Nottingham, NG7 2RD, UK}
\affiliation{Centre for the Mathematics and Theoretical Physics of Quantum Non-equilibrium Systems,
University of Nottingham, Nottingham NG7 2RD, UK}
\author{J. Min\'{a}\v{r}}
\affiliation{School of Physics and Astronomy, University of Nottingham, Nottingham, NG7 2RD, UK}
\affiliation{Centre for the Mathematics and Theoretical Physics of Quantum Non-equilibrium Systems,
University of Nottingham, Nottingham NG7 2RD, UK}
\affiliation{Department of Physics, Lancaster University, Lancaster,  LA1 4YB, UK}
\author{J. P. Garrahan}
\affiliation{School of Physics and Astronomy, University of Nottingham, Nottingham, NG7 2RD, UK}
\affiliation{Centre for the Mathematics and Theoretical Physics of Quantum Non-equilibrium Systems,
University of Nottingham, Nottingham NG7 2RD, UK}
\author{I. Lesanovsky}
\affiliation{School of Physics and Astronomy, University of Nottingham, Nottingham, NG7 2RD, UK}
\affiliation{Centre for the Mathematics and Theoretical Physics of Quantum Non-equilibrium Systems,
University of Nottingham, Nottingham NG7 2RD, UK}
\author{M. Marcuzzi}
\affiliation{School of Physics and Astronomy, University of Nottingham, Nottingham, NG7 2RD, UK}
\affiliation{Centre for the Mathematics and Theoretical Physics of Quantum Non-equilibrium Systems,
University of Nottingham, Nottingham NG7 2RD, UK}

\begin{abstract}
We investigate large deviations of the work performed in a quantum quench across two different phases separated by a quantum critical point, using as example the Dicke model quenched from its superradiant to its normal phase. We extract the distribution of the work from the Loschmidt amplitude and compute for both the corresponding large-deviation forms. Comparing these findings with the predictions of the classification scheme put forward in  [Phys. Rev. Lett. \textbf{109}, 250602 (2012)], we are able to identify a regime which is in fact distinct to the ones identified so far: here the rate function exhibits a non-analytical point which is a strong indication of the existence of an out-of-equilibrium phase transition in the rare fluctuations of the work.
\end{abstract}

\pacs{}
\maketitle

\emph{Introduction---} Understanding out-of-equilibrium phenomena in classical and quantum many-body systems is one of the modern challenges in condensed matter and statistical physics. The breakdown of equilibrium conditions, associated to the absence of detailed balance in the microscopic processes governing the dynamics, results in asymptotic states which do not take the equilibrium Boltzmann-Gibbs form. Some concepts and techniques from thermodynamics and statistical mechanics can be however transferred to an out-of-equilibrium regime, leading for instance to fluctuation-dissipation relations (which connect the response of a system under a weak external perturbation to the correlation between equilibrium thermal fluctuations, see, e.g., \cite{Cugliandolo2002,Crisanti2003}) and fluctuation \cite{Jarzynski2011} relations.

%

In closed quantum systems, the simplest conceptual protocol to obtain an out-of-equilibrium evolution is a \emph{quantum quench}. Physically, this can be thought of as an abrupt change $\Omega_0 \to \Omega$ of one of the external fields appearing in the Hamiltonian $H$, fast enough for the state of the system not to appreciably change across its variation. Typically, one starts from the ground state $\ket{\mathrm{GS} (\Omega_0)}$ of $H(\Omega_0)$ before the quench (time $t  \to 0^-$) and subsequently evolves it for $t>0$ with $H(\Omega)$. Such quantum quenches have been extensively studied to understand relaxation and thermalization in closed quantum systems \cite{Berges2004,Polkovnikov2011} and their relation to integrability, both theoretically \cite{Rigol2007,Calabrese2011} and experimentally \cite{Greiner2002,Kinoshita2006,Gring2012}.

Interestingly, the notion of work can be generalized to the quantum regime and fluctuation relations have been found to hold as well \cite{Esposito2009,Campisi2011}. Furthermore, it has been established that the Loschmidt amplitude $L(t)$ for a quenched system satisfies, in the thermodynamic limit $N\rightarrow \infty$, a large deviation principle $L(t) \sim e^{N l(t)}$. The analytical continuation of $l(t)$ to imaginary time $t\rightarrow -i s$ is related (via a Legendre transform) to the statistics of the work done on the system by the quench \citep{Silva:PRL:2008}. The function $l(-is)$ is typically referred to as \emph{scaled cumulant generating function} (SCGF for short). Gambassi and Silva \cite{Gambassi:PRL:2012} provided a first classification of the possible forms of these large deviation functions, identifying two distinct kinds of qualitative behaviors: for systems in class A (spectrum bounded from above), the SCGF is defined for all values of $s \in \mathbb R$, whereas for systems in class B (spectrum unbounded from above) the SCGF is defined only for values of $s$ larger than a certain threshold value $s > s^*$.

In this work we shed light on the behavior of large fluctuations in the work performed during a quench across a quantum critical point. We show that the statistics of the work may exhibit a non-analytical point, corresponding to a non-equilibrium phase transition, a situation encountered in the studies of the rare events of out-of-equilibrium classical stochastic systems \cite{Carlos:PRE:2013,Carlos:PRL:2017}. Importantly, this constitutes a novel feature of the statistics of the work fluctuations \emph{not included} in the classification scheme put forward in Ref.~\cite{Gambassi:PRL:2012}. For the sake of concreteness we illustrate our ideas using the Dicke model \cite{Dicke:PhysRev:1954}, a paradigmatic Hamiltonian of light-matter interaction. In the past decade, extensive investigations addressed its implementation \cite{Dimer_2007} and connection to the low-energy physics of Bose-Einstein condensates in optical cavities \cite{Nagy:PRL:2010}, its hallmark superradiant phase transition \cite{Lieb:PRA:1973,Emary:PRE:2003} (experimentally probed in \cite{Baumann:Nature:2010,Baumann:PRL:2011,Klinder_2015}), the associated critical phenomena \cite{Gammelmark_2011,Bakemeier_2012,Castanos_2012} and non-equilibrium properties \cite{Bhaseen_2012,Paraan:PRE:2009}, its connection to the physics of spin glasses \cite{Strack:PRL:2011,Buchhold_2013,Rotondo:PRB:2015} and neural networks \cite{Goldbart:PRL:2011,Rotondo:PRL:2015} and its application in the context of the self-organization of the atomic motion \cite{Domokos_2002,Zippilli_2004,Asboth_2005,Black_2003}.

Exploiting the inherent integrability of this model in the thermodynamic limit \cite{Lieb:PRA:1973, Emary:PRE:2003} to construct an explicit, though approximate, representation for the distribution of the work, we highlight a parameter regime going beyond the classification proposed in \cite{Gambassi:PRL:2012} and the corresponding structure of the rate function.
Conceptually, we proceed by: (i) establishing a convenient formalism to describe quenches in the Dicke model; (ii) extracting the large-deviation form of the Loschmidt amplitude in the thermodynamic limit; (iii) highlighting the emergence of a point of non-analyticity in the corresponding rate function describing the fluctuations of the work.

\emph{The model---} We start by setting the notation and recalling the Dicke Hamiltonian in natural units ($\hbar=1$)
\begin{equation}
H_{\mathrm D} (\Omega) = \omega a^{\dagger} a + \Delta J^z +\frac{\Omega}{\sqrt N} (a+a^{\dagger}) J^x\,, 
\label{DH}
\end{equation}
where $a$, $a^\dag$ are bosonic annihilation and creation operators for a single photonic mode of frequency $\omega$. The $J^{\alpha}$'s ($\alpha = x,y,z$) are collective variables describing an ensemble of $N$ spin-$\frac{1}{2}$ atoms which effectively behaves like a single larger spin. These operators satisfy the standard $SU(2)$ commutation relations $\comm{J^\alpha}{J^\beta} = i \epsilon_{\alpha \beta \gamma} J^\gamma$ and we work in the largest representation, where $J^2 = J^\alpha J_\alpha = N(N+2)/4$. The parameter $\Delta$ is the energy cost to ``flip an atomic spin''. The light-matter coupling constant $\Omega$ is divided by $\sqrt{N}$ to ensure that the energy is extensive in the number of atoms.

The Dicke model undergoes a continuous quantum phase transition at $\Omega = \Omega_c = \sqrt{\omega \Delta}$. Below $\Omega_c$ the system is in the normal phase (NP) and the average density of photons $\av{a^\dag a}/N$ in the ground state (GS) vanishes in the thermodynamic limit $N \to \infty$. For $\Omega > \Omega_c$, the system is in the superradiant phase (SP) and develops a macroscopic cavity field, i.e. the average density of photons converges to a finite value. Correspondingly, the average expectations $\av{a + a^\dag}/N$ and $\av{J^x}/N$ also acquire a finite value, resulting in spontaneous breaking of the $\Z_2$ symmetry $U = \rme{i\pi (a^\dag a + J^z)}$. This phase transition has first been studied by Hepp and Lieb \cite{Lieb:AnnPhys:1973, Lieb:PRA:1973,Wang:PRA:1973}, who computed the full partition function of the model in the thermodynamic limit $N \rightarrow \infty$. 


\begin{figure}[t]
\includegraphics[width=0.45 \textwidth]{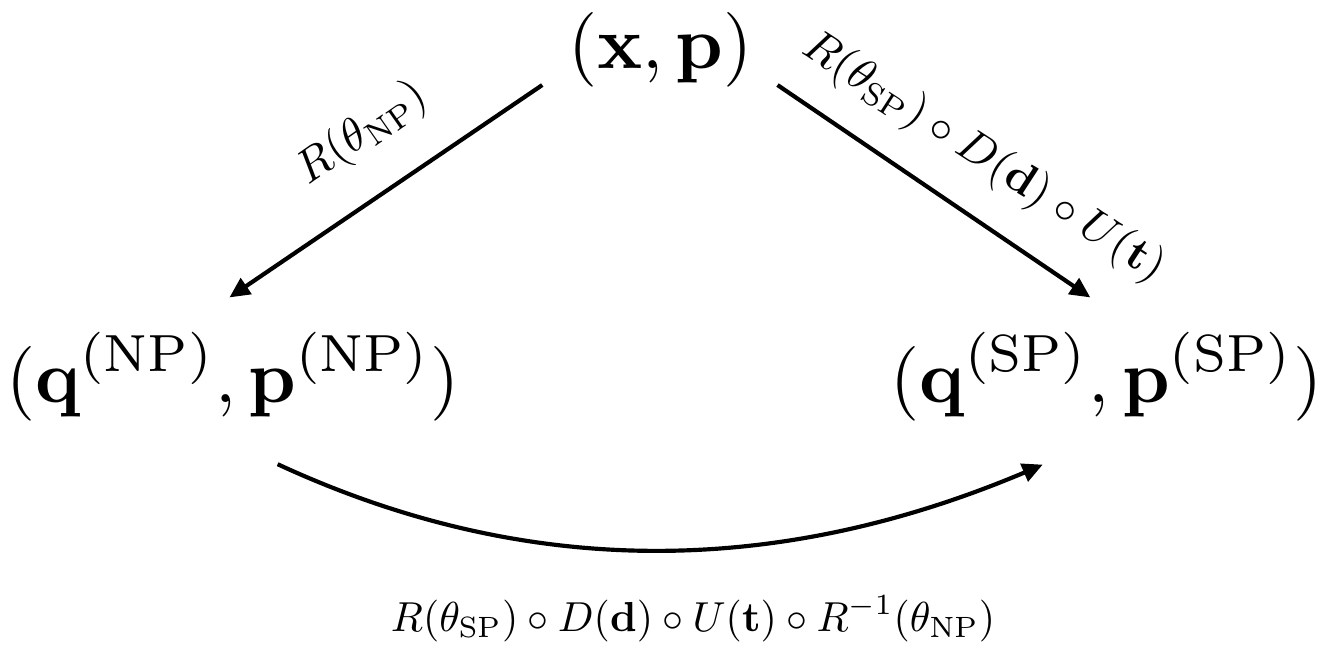}
\caption{Diagram of the coordinate transformations that diagonalize the thermodynamic limit of the Dicke Hamiltonian both in the NP and SP. The original pair of canonical coordinates $(\mathbf{x}, \mathbf{p})$ in Eqs. (\ref{basis1}, \ref{basis2}) are transformed via a rotation $R(\theta_{\textrm{NP}})$ to diagonalize the Hamiltonian in the NP. For the SP the transformation is more involved, requiring to compose a translation represented by $U(\mathbf{t})$, a dilation $D(\mathbf{d})$ and another rotation $R(\theta_{\textrm{SP}})$. The overall transformation that links NP and SP coordinates is easily obtained by composing succesive transformations (see also \cite{SM}).}
\label{Fig1}
\end{figure}

\emph{NP and SP: the coordinate picture---} We work here in the formalism developed in Ref.~\cite{Emary:PRE:2003}, which effectively maps the Dicke Hamiltonian onto a two-boson model. This is achieved via the Holstein-Primakoff transformation $J^z = b^{\dagger} b - N/2$, $\quad J^+ = b^{\dagger} \sqrt{N-b^{\dagger} b}$, $\quad  J^- = \lt  \sqrt{N-b^{\dagger} b} \rt b$, where $b, b^{\dagger}$ satisfy ordinary bosonic commutation relations. By dropping terms proportional to $1/N$, in the NP one obtains a quadratic bosonic Hamiltonian \cite{Emary:PRE:2003, SM}
\begin{equation}
H_{\mathrm{NP}}(\Omega) = \omega a^{\dagger} a +\Delta b^{\dagger} b +\Omega (a+ a^{\dagger})(b+b^{\dagger})- \frac{N \Delta}{2}\,.
\label{eq:H NP}
\end{equation}
In the SP, instead, $b^\dag b$ acquires an extensive component which needs to be singled out in order to correctly truncate the $1/N$ expansion \cite{Emary:PRE:2003}, yielding another quadratic effective Hamiltonian $H_{\textrm{SP}}$ (see \cite{SM} for the explicit expression).

In the thermodynamic limit, $H_{\textrm{NP}}$ and $H_{\textrm{SP}}$ capture, separately in each phase, the thermodynamic properties of the Dicke model. More specifically, one can interpret them as an effective description of the dominating Gaussian fluctuations of the order parameter far from the critical region. As such, this description is rather generic for many-body statistical systems undergoing a discrete symmetry breaking. As shown below, the specific choice of the Dicke model as an example allows us to more easily establish the connection between the relevant fluctuations in the two phases. This description is valid up to corrections which become irrelevant when increasing the system size. 

Hereafter, we neglect the $1/N$ corrections in $H_{\textrm{NP/SP}}$, which is equivalent to taking the thermodynamic limit $N \to \infty$ before calculating the time evolution of the system. With this approximation, the quench is mapped onto the non-equilibrium dynamics of a two-dimensional harmonic oscillator (2DHO) and can be exactly solved to highlight the non-analyticities of the rate function.


Since $H_{\rm NP}$ and $H_{\rm SP}$ are quadratic, they can be diagonalized via appropriate generalized Bogolyubov transformations \cite{Bogolyubov1947}. However, for our purposes it is more convenient to work in the aforementioned 2DHO representation and consider the associated coordinate representation $\mathbf{x} = (x,y)$, $\mathbf{p} = (p_x, p_y)$:
\begin{align}
x &= \frac{1}{\sqrt{2 \omega}} (a + a^{\dagger})\,, \quad p_x = i \sqrt{\frac{\omega}{2}} (a^{\dagger} - a), 
\label{basis1}
\\ 
y &= \frac{1}{\sqrt{2 \Delta}} (b + b^{\dagger})\,, \quad p_y = i \sqrt{\frac{\Delta}{2}} (b^{\dagger} - b)\,,
\label{basis2}
\end{align}
where $\comm{x}{p_x} = \comm{y}{p_y} = i$ and all other commutators vanish. In each phase, the Bogolyubov transformation which diagonalizes $H_{\mathrm{NP} / \mathrm{SP}}$ becomes a geometric transformation of the coordinates, namely a combination of rotations $R$, dilations $D$ and translations $U$ as shown in Fig.~\ref{Fig1}. In the diagonal basis, these are written in terms of new coordinates $\mathbf{q}^{(\nu)} = (q_x^{(\nu)}, q_y^{(\nu)})$, $\mathbf{p}^{(\nu)} = (p_x^{(\nu)}, p_y^{(\nu)})$ and have fundamental frequencies $\omega_{\pm}^{(\nu)}$, where $\nu = \mathrm{NP}, \mathrm{SP}$ \cite{SM}. 

\emph{Loschmidt amplitude---} A quantity that is fundamental for characterizing the work statistics is the Loschmidt amplitude~\cite{Silva:PRL:2008}
\begin{equation}
L(t) = \bra{\mathrm{GS} (\Omega_0)} e^{-i t H_{\rm NP}(\Omega)} \ket{\mathrm{GS} (\Omega_0)},
\label{Loschmidt}
\end{equation}
with $\ket{\mathrm{GS} (\Omega_0)}$ denoting one of the two superradiant ground states at maximal transverse magnetization, i.e.~chosen as the ground state of $H_{\rm SP} (\Omega_0) + \epsilon J^x$ for $\epsilon \to 0^+$ (the other one would be obtained by minimizing the energy for $\epsilon \to 0^-$, corresponding to $J^x \to - J^x$). Without loss of generality, we rescale the energies so that the NP ground state has zero energy. 

Inserting four completeness relations in (\ref{Loschmidt}), the Loschmidt amplitude becomes
\begin{widetext}
\begin{align}
L(t) =& \int d^2 q_1^{(\mathrm{SP})} d^2 q_2^{(\mathrm{SP})} d^2 q_1^{(\mathrm{NP})} d^2 q_2^{(\mathrm{NP})} \braket{\mathrm{GS}(\Omega_0)| \mathbf{q}_1^{(\mathrm{SP})}} \braket{\mathbf{q}_1^{(\mathrm{SP})}|\mathbf{q}_1^{(\mathrm{NP})}} \nonumber \\
&\bra{\mathbf{q}_1^{(\mathrm{NP})}}e^{-i t H_{\mathrm{NP}}} \ket{\mathbf{q}_2^{(\mathrm{NP})}} \braket{\mathbf{q}_2^{(\mathrm{NP})}|\mathbf{q}_2^{(\mathrm{SP})}} \braket{\mathbf{q}_2^{(\mathrm{SP})} |\mathrm{GS}(\Omega_0)}\,,
\label{Losch2}
\end{align}
\end{widetext} 
In the expression above, we note that: (i) $\braket{\mathrm{GS}(\Omega_0) | \mathbf{q}_1^{(\mathrm{SP})}}$ and $\braket{\mathbf{q}_2^{(\mathrm{SP})} |\mathrm{GS(\Omega_0)} }$ are the ground state wavefunctions of the SP two-dimensional harmonic oscillator and are therefore (as functions of $\mathbf{q}_{1/2}^{(\mathrm{SP})}$) Gaussians with zero mean and variances $ \left(1/\sqrt{\omega_{+}^{(\mathrm{SP})}},1/\sqrt{\omega_{-}^{(\mathrm{SP})}}\right)$; (ii) $\bra{\mathbf{q}_1^{(\mathrm{NP})}}e^{-i t H_{\mathrm{NP}}} \ket{\mathbf{q}_2^{(\mathrm{NP})}}$ is the propagator of the NP two-dimensional oscillator, and thus has a complex Gaussian structure which becomes purely Gaussian after a Wick rotation to imaginary time $t \to -is$; (iii) the overlaps $ \braket{\mathbf{q}_j^{(\mathrm{NP})}|\mathbf{q}_j^{(\mathrm{SP})}}$ correspond to a change of variable \cite{SM} in the integration according to the canonical transformation mapping $\mathbf{q}^{(\mathrm{NP})} \leftrightarrow \mathbf{q}^{(\mathrm{SP})}$ (see Fig.~\ref{Fig1}), which can be expressed as
\begin{equation}
 \mathbf{q}^{(\mathrm{SP})} = S \mathbf{q}^{(\mathrm{NP})}   + \sqrt{N}\mathbf{T}\,,
\end{equation}
where we introduced the shorthand $S = R (\theta_{\mathrm{SP}})D(\mathbf d)R^{-1} (\theta_{\mathrm {NP}})$ and $\sqrt{N} \mathbf{T} = R (\theta_{\mathrm{SP}}) D(\mathbf d)\mathbf t$ in relation to the sketch in Fig.~\ref{Fig1}. The exact expressions of the geometric parameters $(\theta_{\mathrm{SP/NP}}, \mathbf d, \mathbf t)$ in terms of the physical ones $(\Omega, \Omega_0, \omega, \Delta)$ can be found in \cite{SM}. The problem of calculating $L(t)$ is now reduced to a Gaussian integration, which can be solved exactly to yield a large deviation form
\begin{equation}
L(t) = A(t) e^{N l(t)}\,,
\label{LDLosch}
\end{equation}   
where both the function $l(t)$ and the prefactor $A(t)$ are intensive functions, i.e. do not depend on $N$. To write them in a compact form, we introduce three diagonal matrices $Q_{\mathrm{SP}}  = \mathrm{diag} \left(\omega_{+}^{(\mathrm{SP})},\omega_{-}^{(\mathrm{SP})} \right)$, $P_\pm (t) = \pm i \, \mathrm{diag} \left(\omega_{+}^{(\mathrm{NP})} \lt \tan \lt \frac{ \omega_{+}^{(\mathrm{NP}) } t}{2} \rt   \rt^{\pm 1},\omega_{-}^{(\mathrm{NP})} \lt \tan \lt \frac{ \omega_{-}^{(\mathrm{NP}) } t}{2} \rt   \rt^{\pm 1} \right)$ and the (generally) non-diagonal ones
\begin{equation}
\mathcal K_{\pm} (t) =  S^\intercal Q_{\mathrm{SP}} S + P_\pm(t) .
\label{detK}
\end{equation}
In terms of these matrices, the rate function reads
\begin{equation}
l(t) = -{\mathbf T}^\intercal \left(Q_{\mathrm{SP}} - Q_{\mathrm{SP}} S \mathcal K_+^{-1}(t) S^\intercal Q_{\mathrm{SP}}\right)  {\mathbf T}\,,
\label{SCGF}
\end{equation}
\begin{figure*}[t]
\includegraphics[width=0.9 \textwidth]{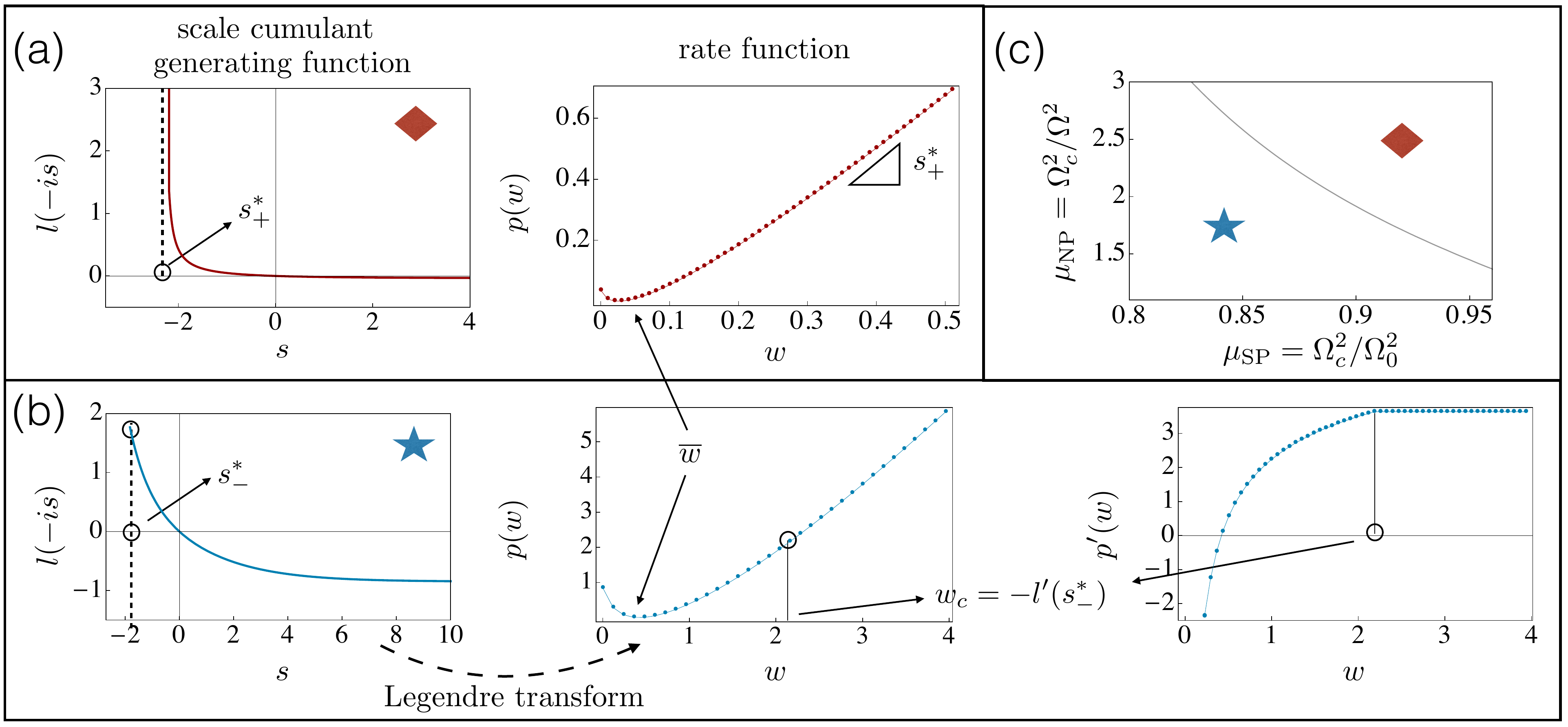}
\caption{Phase transitions in the work distribution (at fixed $\omega=1$). (a) The SCGF $l(-i s)$ diverges as $s \rightarrow s^*_+$. Correspondingly, $p(w)$, which is obtained by a Legendre transform (LT) from $l(-i s)$, approaches asymptotically the linear regime with slope $-s^*_+$ for $w \gg \overline w$ (with $\overline w$ the typical value of the work, i.e. the one for which $p(\overline w) = 0$). This scenario is the one expected for the models that belong to the class B introduced in Ref. \cite{Gambassi:PRL:2012}. (b) The singular behavior of $l(- i s)$ is controlled by $s^*_-$ which is approached at finite value with finite derivative $l'(s^*_-)$. The LT thus displays a non-analytical point in $w_c= -l'(- i s^*_-)$ and $p(w) = -s^*_- w$ for $w > w_c$. This non-analytical point of the rate function corresponds to a phase transition in the rare fluctuations of the work. Plots in (a) and (b) are obtained with the same $\Delta = 0.8$ and $\Omega = 0.3$ by tuning the superradiant coupling from $\Omega_0 = 0.47$ to $\Omega_0 = 0.9$. (c) Analytical (diamond) and non-analytical (star) domains of $p(w)$ in the plane $\mu_{\textrm{NP}}$-$\mu_{\textrm{SP}}$ with $\mu_{\textrm{NP}} = \Omega_c^2/\Omega^2$ and $\mu_{\textrm{SP}} = \Omega_c^2/\Omega_0^2$ at $\Delta = 0.8$. At fixed $\Omega$, the non-analytical point appears only if the quench starts deep enough in the SP. The solid line denotes the boundary between the two areas and is where the non-analyticity in $p(w)$ appears. }
\label{Fig2}
\end{figure*}
while the prefactor is
\begin{equation}
A(t) = \sqrt{\frac{-4 \det D^2 \,  \omega_+^{(\mathrm{NP})}  \omega_-^{(\mathrm{NP})}  \omega_+^{(\mathrm{SP})}  \omega_-^{(\mathrm{SP})} }{\sin \left( \omega_+^{(\mathrm{NP})} t\right) \sin \left( \omega_-^{(\mathrm{NP})} t\right) \det \mathcal{K}_+(t) \det \mathcal{K}_-(t)}}\,.
\label{eq:At}
\end{equation}

\emph{Statistics of work---} The Loschmidt amplitude calculated above gives access to the statistics of the work done by the quench $\Omega_0 \rightarrow \Omega$, as shown in \cite{Gambassi:PRL:2012}. The average work per atom $w= W_N/N$ is a stochastic variable with a distribution $P(w)$ whose generating function is the analytical continuation of $L(t)$ to imaginary time $t \rightarrow -i s$. In the large-$N$ limit, $L(-is)$ can be written as in Eq.~\eqref{LDLosch} by substitution. By the Gartner-Ellis theorem \cite{Touchette:PhysRep:2009}, $P(w)$ fulfills a large deviation principle as well, namely $P(w) \propto \exp(-N p(w))$ and furthermore the rate function $p(w)$ and the SCGF $l(-is)$ are related by a Legendre transform $p(w) = -\inf_{s\in \mathbb R} (w s + l(-i s))$. In the following we investigate the behavior of the rate function $p(w)$ starting from the SCGF $l(-i s)$ in Eq. \eqref{SCGF}.

The Dicke model has no upper bound to its energy spectrum and thus should belong to class B, according to Ref. \cite{Gambassi:PRL:2012}. However, in some parameter regimes its SCGF $l(-is)$ reaches the leftmost point of its domain with finite (instead of diverging) derivative, see Fig.~\ref{Fig2}(b). This property of $l(-i s)$ translates in a non-analytical behavior of $p(w)$, which was not previoulsy reported. 

In order to visualize the appearance of such a regime we need to study the domain of $L(-is) = \bra{\mathrm{GS} (\Omega_0)} e^{- s H_{\rm NP}(\Omega)} \ket{\mathrm{GS} (\Omega_0)}$. In performing the analytic continuation of \eqref{LDLosch}, we thereby have to stop at the first singularity encountered. First, we note that for $s\geq 0$ $L(-is)$ is always well-defined. Second, a singularity of $L(-is)$ can be either associated to a singularity of SCGF $l(-is)$ or of the amplitude $A(-is)$ (or both). Third, for $l(-is)$ this can only occur when $\det \mathcal{K}_+(-is) = 0$, whereas for $A(-is)$ singularities can additionally emerge when $\det \mathcal{K}_-(-is) = 0$. We denote by $s^\ast_\pm$ the rightmost singular point of $\mathcal{K}_\pm(-is)$. For completeness, we remark that $\sin \left( \omega_+^{(\mathrm{NP})} t\right) \sin \left( \omega_-^{(\mathrm{NP})} t\right) \det \mathcal{K}_-(t) \stackrel{t\to 0}{\to}  \omega_-^{(\mathrm{NP})}  \omega_+^{(\mathrm{NP})} / 4$, curing the singularity in $s=0$ and implying $s^\ast_- < 0$. The left domain edge of $L(-is)$ is therefore $\max(s^\ast_+, s^\ast_-)$.
Two regimes can thus be identified: if $s^*_+ > s^*_-$, the SCGF $l(-is)$ diverges at its leftmost point (corresponding to class B of Ref.~\cite{Gambassi:PRL:2012}). Correspondingly, $p(w)$ approaches asymptotically a linear regime with slope $-s^*_+$ for $w \to +\infty$, as sketched in Fig.~\ref{Fig2}(a). If instead $s^*_+ < s^*_-$, then $l(-is)$ remains finite and differentiable in $s^*_-$, signalling a different qualitative behavior. This in fact yields a point of non-analyticity in $p(w)$ located at $w_c = -l'(-i s^*_-) < +\infty$, sketched in Fig.~\ref{Fig2}(b). For all $w > w_c$, $p(w) = -s^*_- w$ exactly, corresponding to a jump in the second derivative of $p(w)$, which vanishes for $w > w_c$, see Fig. \ref{Fig2}(b).
The emergence of either scenario depends on the pre- and post-quench parameters $\Omega_0$, $\Omega$, $\omega$ and $\Delta$. In Fig.~\ref{Fig2}(c) we show a numerical determination of these two regimes for $\Delta = 0.8$, $\omega=1$ and various values of $\Omega_0$ and $\Omega$.    
A non-analytical point in the rate function $p(w)$ signals a phase transition (see e.g. \cite{Znidaric:PRL:2014,Manzano:PRB:2014,Carollo:PRE:2017}) occurring, in our case, in the statistics of the rare fluctuations of the work done. In a classical context, this would correspond to a macroscopic change in the nature of the ``typical'' configurations the system would display at \emph{fixed} average work. By analogy, it seems reasonable to expect that in a quantum context this transition will correspond to an abrupt change in the post-quench eigenvectors which contribute the most to expectations at fixed average energy. These aspects go however beyond the scope of the present work and will be addressed in detail at a later stage.

\emph{Discussion and Conclusions---} In conclusion, we identified singularities in the distribution function of work for a system undergoing a quantum quench, extending the original classification presented in Ref.~\cite{Gambassi:PRL:2012}.
In particular, the rate function describing the statistics of the work exhibits in this regime a non-analytical point, signaling an out-of-equilibrium phase transition in the rare fluctuations of the work. Non-analiticities in the rate functions describing rare events are well-established in the context of classical stochastic processes out of equilibrium \cite{Znidaric:PRL:2014,Manzano:PRB:2014,Carollo:PRE:2017,Kundu:JStat:2011,Sanjib:PRE:2012} and have been identified in the counting statistics of continuously-measured quantum systems \cite{Genway2012,Hickey2013,Hickey2014,Hickey2014_2} and in the large-deviations of diffusing cold atoms \cite{Barkai:PRL:2017}. This manuscript highlights the emergence of this scenario in quenched closed quantum systems. It would be also interesting to understand what relation, if any, there is between the non-analiticities found here and the dynamical phase transitions investigated in \cite{Heyl:PRL:2013}.

\emph{Acknowledgements---} The research leading to these results has received funding from the European Research Council  under  the  European  Unions  Seventh  Framework  Programme  (FP/2007-2013)/ERC  Grant  Agreement  No. 335266  (ESCQUMA). P.R. acknowledges funding by the European Union through the H2020 - MCIF No. $766442$. I.L. gratefully acknowledges funding  through  the  Royal  Society  Wolfson  Research Merit Award. P.R. acknowledges funding by the European Union through the H2020 - MCIF No. 766442. The authors wish to thank M. Heyl and A. Gambassi for useful discussions at a preliminary stage of the project and H. Touchette for useful comments on the manuscript.

\begin{widetext}
\section{Appendix: Dicke model and Holstein-Primakoff transformation}

The exact form in the thermodynamic limit of the ground state of the Dicke model can be worked out both in the normal (NP) and in the superradiant phase (SP) through a suitable Holstein-Primakoff transformation. Let us consider the Dicke Hamiltonian:
\begin{equation}
H_{\mathrm D} = \omega a^{\dagger} a + \Delta J^z +\frac{\Omega}{\sqrt N} (a+a^{\dagger}) J^x\,, 
\end{equation}
where the $J^{\alpha}$'s ($\alpha = x,y,z$) form an irreducible representation of the angular momentum of dimension $N/2$. In the large-$N$ limit we can employ the following Holstein-Primakoff transformation:
\begin{equation}
J^z = b^{\dagger} b - N/2\,, \quad J^+ = b^{\dagger} \sqrt{N-b^{\dagger} b}\,, \quad  J^- =  \sqrt{N-b^{\dagger} b} \,b\,,
\end{equation} 
where $b, b^{\dagger}$ satisfy ordinary bosonic commutation relations. This leads to:
\begin{equation}
H_{\mathrm D} = \omega a^{\dagger} a + \Delta b^{\dagger} b + \Omega (a+a^{\dagger})\left(b^{\dagger} \sqrt{1-\frac{b^{\dagger} b}{N}} + \sqrt{1-\frac{b^{\dagger} b}{N}} \,b\right) - \frac{N \Delta}{2}\,.
\label{DickeHP}
\end{equation}
In the thermodynamic limit we can naively ignore the terms proportional to $1/\sqrt N$. In this way we obtain a solvable quadratic bosonic model. This works in the NP at small $\Omega$, but this approximation breaks down in the SP for $\Omega$ large enough, signaling that a quantum phase transition is taking place. In the following we analyze separately the two different phases obtaining the corresponding effective Hamiltonians. We will focus, in particular, on the coordinate representation, that will be useful to evaluate the Loschmidt amplitudes. The results summarized here in the following two subsections are extensively covered in \cite{Emary:PRE:2003}.

\subsection{Normal phase}

In the NP we omit the $1/\sqrt N$ terms so that Eq. \eqref{DickeHP} reduces to
\begin{equation}
H_{\mathrm{NP}} = \omega a^{\dagger} a +\Delta b^{\dagger} b +\Omega (a+ a^{\dagger})(b+b^{\dagger})- \frac{N \Delta}{2}\,.
\end{equation}
This Hamiltonian can be diagonalized by a suitable Bogoliubov transformation that mixes the four different creation and annihilation operators. However the picture is simpler if we switch to the coordinate space, by writing:
\begin{equation}
x = \frac{1}{\sqrt{2 \omega}} (a + a^{\dagger})\,, \quad p_x = i \sqrt{\frac{\omega}{2}} (a^{\dagger} - a), \quad y = \frac{1}{\sqrt{2 \Delta}} (b + b^{\dagger})\,, \quad p_y = i \sqrt{\frac{\Delta}{2}} (b^{\dagger} - b)\,.
\label{basis}
\end{equation}
In this way, it is easy to realize that a rotation in the $(x,y)$-plane puts the Hamiltonian in a diagonal form. In particular we need the following coordinate transformation:
\begin{equation}
\mathbf{q}^{(\mathrm{NP})} = (q_x^{(\mathrm{NP})}, q_y^{(\mathrm{NP})})^T = R (\theta_{\mathrm {NP}}) \mathbf x,\quad R (\theta) = \left( \begin{array}{ccc}
\cos(\theta) & \sin(\theta)  \\
-\sin(\theta) & \cos(\theta)  \end{array} \right) 
\label{coordnp}
\end{equation} 
with $\theta_{\mathrm{NP}}$ given by:
\begin{equation}
\tan (2 \theta_{\mathrm{NP}}) = \frac{4 \Omega \sqrt{\omega \Delta}}{\omega^2-\Delta^2}\,.
\end{equation}
The eigenfrequencies of the Hamiltonian in the NP read:
\begin{equation}
\omega_{\pm}^{(\mathrm{NP})} = \sqrt{\frac{1}{2} \left(\omega^2 + \Delta^2 \pm \sqrt{(\omega^2-\Delta^2)^2 + 16 \Omega^2 \omega \Delta} \right)}\,.
\label{eigenf}
\end{equation}    
It follows from Eq. \eqref{eigenf} that the potential is not bounded from below for $\Omega > \sqrt{\omega \Delta}/2 = \Omega_c$, thus signaling that this effective Hamiltonian description breaks down at strong coupling. Before moving to the analysis of the SP, we notice that the ground state of the NP in the coordinate basis $\mathbf{q^{(\mathrm{NP})}}$ is a $2$-dimensional Gaussian centered around $\mathbf{q^{(\mathrm{NP})}} = (0,0)$ with variance $(\sigma_x,\sigma_y) = \left(1/\sqrt{\omega_{+}^{(\mathrm{NP})}},1/\sqrt{\omega_{-}^{(\mathrm{NP})}}\right)$.

\subsection{Superradiant phase}

The derivation of the effective Hamiltonian in the SP is more involved. We refer to \cite{Emary:PRE:2003} for the details. Here we only report the fundamental results that we will need in the following to compute the Loschmidt amplitude corresponding to the quench for the NP to the SP. Starting from Eq. \eqref{DickeHP} we define  define two new operators $c = a -\sqrt \alpha$, $d = b + \sqrt \beta$ and we choose the displacement parameters properly in order to eliminate the terms linear in the bosonic operators. In this way we get an effective Hamiltonian for the SP, that reads:
\begin{align}
H_{\textrm{SP}} &= \omega c^{\dagger} c + \frac{\Delta (1+ \mu)}{2\mu} d^{\dagger} d + \frac{\Delta (1-\mu)(3+\mu)}{8\mu(1+\mu)} (d^{\dagger} + d)^2 \\ \nonumber
&+\Omega \mu \sqrt{\frac{2}{1+\mu}} (c^{\dagger} + c)(d^{\dagger} + d)- \frac{N}{2} \left(\frac{2\Omega^2}{\omega} + \frac{\Delta^2 \omega}{8\Omega^2}\right) - \frac{\Omega^2 (1-\mu)}{\omega}\,,
\end{align}
where $\mu = (\omega \Delta)/(4 \Omega^2) =  (\Omega_c/\Omega)^2$. Again we focus on the coordinate space representation. In order to get the diagonal form of the hamiltonian in the SP, we need to apply three succesive canonical transformations, firstly a translation represented by the vector $\mathbf{t} = (t_x, t_y)$ (which is the transformation that allows to get $H_{\textrm{SP}}$ from Eq. \eqref{DickeHP}), then a dilation $D(\mathbf{d})$ on the $y$ coordinate and finally a rotation by an angle $\theta_{\mathrm{SP}}$. In formulas:
\begin{equation}
\mathbf{q}^{(\mathrm{SP})} = R (\theta_{\mathrm{SP}}) D(\mathbf{d}) (\mathbf x + \mathbf t)\,,
\label{coordsp}
\end{equation}
where the explicit parameters of the transformation are:
\begin{align}
&\mathbf t = \left(\frac{\sqrt 2 \Omega}{\omega} \sqrt{\frac{N (1-\mu^2)}{\omega}},\sqrt{\frac{N(1-\mu)}{\Delta}} \right)\,, \\
&\tan (2 \theta_{\mathrm{SP}}) = \frac{2 \omega \Delta \mu^2}{\mu^2\omega^2-\Delta^2}\,, \\
&\mathbf{d} = \left(1, \sqrt{\frac{2\mu}{1+\mu}}\right)\,.
\end{align}

In the new coordinates of Eq. \eqref{coordsp} the effective Hamiltonian $H_{\mathrm{SP}}$ is diagonal with eigenfrequencies given by:
\begin{equation}
\omega_{\pm}^{(\mathrm{SP})} = \sqrt{\frac{1}{2} \left(\omega^2 + \frac{\Delta^2}{\mu^2} \pm \sqrt{\left(\omega^2-\frac{\Delta^2}{\mu^2}\right)^2 + 4 \omega^2 \Delta^2} \right)}\,.
\end{equation}
Again, in the coordinate basis $\mathbf q^{(\mathrm{SP})}$ the ground state of the SP is a Gaussian centered around $\mathbf q^{(\mathrm{SP})}= (0,0)$ and with variances $(\sigma_x, \sigma_y) = \left(1/\sqrt{\omega_{+}^{(\mathrm{SP})}},1/\sqrt{\omega_{-}^{(\mathrm{SP})}}\right)$.

Finally, combining Eqs. \ref{coordnp} and \ref{coordsp}, we obtain the explicit relation between NP and SP coordinates that we extensively use in the main text:
\begin{equation}
 \mathbf{q}^{(\mathrm{SP})} = S \mathbf{q}^{(\mathrm{NP})}   + \sqrt{N}\mathbf{T}\,, \quad S = R (\theta_{\mathrm{SP}})D(\mathbf d)R (\theta_{\mathrm {NP}})^\intercal\,, \quad \sqrt{N} \mathbf{T} = R (\theta_{\mathrm{SP}}) D(\mathbf d)\mathbf t\,.
\end{equation}

\end{widetext}

\bibliographystyle{apsrev4-1}

\bibliography{bibquenchdicke}

\end{document}